\documentclass[aps,prd,twocolumn,eqsecnum]{revtex4}
\begin{document}
\title{The electron thermal propagator at $p\gg T$: An entire function of
$p_{0}$}
\author{H.Arthur Weldon}
\affiliation{Department of Physics, West Virginia University, Morgantown,
West Virginia 26506-6315}
\date{September 13, 2003}

\begin{abstract}
The retarded electron propagator
$S_{R}(p_{0},\mathbf{p})$ at high momentum $p\gg T$ was shown by Blaizot and Iancu to
be an entire function of complex $p_{0}$. 
In this paper a specific  form for $S_{R}(p_{0},\textbf{p})$ is obtained
and checked by showing that  its temporal Fourier transform
$S_{R}(t,\mathbf{p})$  has the correct behavior at large $t$.
Potential infrared and collinear divergences from the emission of soft
photons do not occur.
\end{abstract}

\maketitle

\section{Introduction}

High temperature QCD is, in one respect, simpler than high temperature
QED.  In both theories the  damping rates for charged particles at very high
momentum,
$p\gg T$,  have infrared singularities at the mass shell when computed in
perturbation theory using Braaten-Pisarski resummation \cite{RP}. The
divergence comes from the emission and absorption of quasistatic, transverse
gauge bosons.
  In QCD this apparent infrared divergence in the quark and gluon damping
rates   is cut off by the nonperturbative magnetic screening mass for transverse
gluons and this gives finite damping rates
\cite{quark}.  
For  quarks the damping rate $\gamma$  determines the location of the  pole in
the retarded propagator
$s_{R}(p_{0},p)$:
\begin{displaymath}
s_{R}^{\text{quark}}(p_{0},p)={1\over p_{0}-p+i\gamma/2}.\label{quark}
\end{displaymath} 
The Fourier transform gives  the time-dependent retarded propagator 
\begin{displaymath}
s_{R}^{\text{quark}}(t,p)=-i\theta(t)e^{-ipt}\exp[-\gamma t/2]
\end{displaymath}

In QED there is no magnetic screening mass and consequently the lowest
order electron damping rate is infrared divergent 
  \cite{RP,d1,d2,d3,d4}. 
A direct analysis of  the time dependent retarded propagator
$s_{R}(t,p)$   produced surprising results: the time dependent propagator
is not infrared divergent and at large
times, $\omega_{p}t\gg 1$, it falls more rapidly than an exponential 
\begin{equation}
s_{R}(t,p)\to -ie^{-ipt}\exp\Big\{-\alpha Tt\big[\ln(\omega_{p}t)+C\big]\Big\}.
\label{BI}\end{equation}
Here $\alpha =g^{2}/4\pi$ is the fine structure constant and $\omega_{p}=gT/3$ is
the plasma frequency. (It is customary to use $g$ for the electric charge so as
not to confuse it with the base of the natural logarithm $e$.)
Blaizot and Iancu \cite{B1,B2,B3} obtained this result using a functional integral
version of the Bloch-Nordsieck approximation. This resums an infinite class of
Feynman graphs with no fermion loops, except for the Braaten-Pisarksi
hard-thermal-loop correction to the quasistatic transverse photon propagator. 
Wang, Boyanovsky,  de Vega, and  Lee
obtained the same result by using the dynamical renormalization group to improve the
lowest order result \cite{DB1}.  
Boyanovsky et at \cite {DB2} found the same behavior
for the  retarded propagator of the scalar particle in scalar QED. 
Because $s_{R}(t,p)$ falls faster than an exponential, its  Fourier transform
$s_{R}(p_{0},p)$ has some very unusual properties. 
Blaizot and Iancu proved the following about $s_{R}(p_{0},p)$:
(1) It is an entire function of $p_{0}$; (2) It vanishes in the limit
$\text{Im}\,p_{0}\to +\infty$; (3) It diverges in the limit
$\text{Im}\,p_{0}\to -\infty$. Specifically, at $p_{0}=p-i\zeta$ it is pure
imaginary and bounded from below:
\begin{equation}
\zeta\to\infty:\hskip0.5cm 
\big|s_{R}(p\!-\!i\zeta,p)\big| >{1\over \alpha T}
\exp\Big(a\exp\Big[{\zeta\over\alpha T}\Big]\Big).\label{bound}
\end{equation}
This asymptotic behavior was deduced in \cite{B2} using $C=0$ and the constant $a$
was shown to have the value $a=3g\exp(-1)/4\pi $. However, revising the argument of
\cite{B2} for a non-zero value of $C$ gives
\begin{equation}
a= {3g\over 4\pi}\exp[-C-1]\label{a}
\end{equation}
A propagator that is an entire function of  $p_{0}$ has  no poles, no branch
points, and no essential singularities.

 The purpose of this paper is to find the
function
$s_{R}(p_{0},p)$ whose Fourier transform has the asymptotic behavior in Eq.
(\ref{BI}). The physical approximations which lead to Eq. (\ref{BI}) are
valid in  the region
$p_{0}\approx p$ \cite{B1,B2,B3,DB1,DB2} and it is only this region that
is investigated here. 
Sec. II shows that the Bloch-Nordsieck propagator $s_{R}(t,p)$
has the very unusual property of being an entire function of complex $t$.
In Sec. III only the leading term in the asymptotic expansion is employed, viz.
\begin{equation}
-i{\sqrt{2\pi a}\over\alpha T}\exp\bigg[i\,{p_{0}\!-\!p\over 2\alpha T}
+ a\exp\Big[i\,{p_{0}\!-\!p\over \alpha T}\Big]\bigg].\label{ansatz1}
\end{equation} 
 This approximation for $s_{R}(p_{0},p)$ works surprisingly well. It  is checked by
showing that the Fourier transform over energies  $p_{0}\approx p$ 
agrees with Eq. (\ref{BI}) at large $t$.
The motivation for  the ansatz, Eq. (\ref{ansatz1}), is given in Appendix B.
There a complete asymptotic expansion of  $s_{R}(p_{0},p)$ is obtained.
The first term of this expansion is Eq. (\ref{ansatz1}).  
Sec. IV  shows how the absence of singularities in the electron propagator
imply that in hard scattering processes, there will be no infrared or collinear
divergences and the quantitative contribution of these regions is numerically small.
Sec. V discusses some further consequences.

\section{Comment on complex time} 

The logarithmic dependence on $t$ in  Eq. (\ref{BI}) does not imply that
there is a branch cut for negative $t$. 
 The full 
 Bloch-Nordsieck calculation of  Blaizot and Iancu \cite{B1,B2,B3} gives
\begin{displaymath}
s_{R}(t,p)=-i\theta(t) e^{-ipt}\exp\big[-\alpha T t
\,f(\omega_{p}t)\big],
\end{displaymath}
where $f(z)$ is given by
\begin{displaymath}
 f(z)=C-\gamma+{1-e^{-z}\over z}
+\int_{0}^{z}\!\!ds\,{1-e^{-s}\over s}.
\end{displaymath}
This representation shows that $f(z)$ is analytic everywhere in the complex
$z$ plane. (See also Appendix B.) Thus the retarded propagator with $\theta (t)$ omitted
is  analytic everywhere in the complex $t$ plane.

Analyticity in $t$ is very unusual. Appendix A shows that in general the
retarded propagator (without the $\theta(t)$) is the difference of two
functions: one, $S_{>}(t,p)$, is analytic in the open strip
$-\beta<\text{Im}
\,(t) <0$; the 
other   $S_{<}(t,p)$ is analytic in the open strip $0<\text{Im}
\,(t) <\beta$. Thus generally, the retarded
propagator is only defined on the real $t$ axis and is not analytic
off-axis. 
Appendix A also shows that the analyticity
in  $t$ of $s_{R}(t,p)$ does not improve on, or detract from, the validity of
the Kubo-Martin-Schwinger condition \cite{KMS}, which requires 
$\mathcal{S}_{>}(t-i\beta,p
)=-\mathcal{S}_{<}(t,p)$.

\section{ The  propagator for $\mathbf{p_{0}\approx p}$}

 This section will provide evidence that a good
approximation to the retarded electron propagator in the region 
\begin{equation}p-{\pi\alpha
T\over 2}<\text{Re}\,p_{0}<p+{\pi\alpha T\over 2},\label{strip1}
\end{equation}
 is the function
\begin{equation}
s_{R}(p_{0},p)={-}iN\exp\bigg[i\,{p_{0}\!-\!p\over 2\alpha T}
+ a\exp\Big[i\,{p_{0}\!-\!p\over \alpha T}\Big]\!\bigg],\label{ansatz}
\end{equation}
where
$N=\sqrt{2\pi a}/\alpha T$.  It is and entire function of $p_{0}$
and has the correct behavior anticipated in Eq. (\ref{bound}).  The Fourier
transform of Eq. (\ref{ansatz}) will be shown to reproduce Eq. (\ref{BI}) at
large time. The motivation for the ansatz (\ref{ansatz}) is contained in Appendix B. 
That appendix contains a complete asymptotic expansion of $s_{R}(p_{0},p)$
valid in the strip (\ref{strip1}). 
Eq. (\ref{ansatz}) is
first term in the expansion. 

For any propagator, the spectral function (for real $p_{0}$) is  
\begin{displaymath}
\rho(p_{0},p)=i[s_{R}(p_{0},p)-s_{R}(p_{0},p)^{*}].\end{displaymath}
For the approximate propagator Eq. (\ref{ansatz}),
\begin{eqnarray*}
\rho(p_{0},p)
&=&2N\exp\bigg\{a\cos\bigg[{p_{0}\!-\!p\over 2\alpha T}\bigg]\bigg\}\\
&&\times \cos\bigg[{p_{0}\!-\!p\over 2\alpha T}
+ a\sin\Big[{p_{0}\!-\!p\over \alpha T}\Big]\!\bigg].
\end{eqnarray*}
If $a$ is small, this is positive in the range given by Eq. (\ref{strip1}).
For $|p_{0}-p|\ll \alpha\pi T$ the approximate behavior is
\begin{displaymath}
\rho(p_{0},p)\approx 2Ne^{a}\bigg\{1-\Big({1\over 8}+a+{a^{2}\over 2}\Big)
\Big[{p_{0}\!-\!p\over \alpha T}\Big]^{2}\bigg\}.
\end{displaymath}

\subsection{Context for application}

When the fermions are massless in  QCD or QED, 
  chirality
and TCP invariance guarantees that
the matrix structure of the retarded fermion propagator is
\begin{eqnarray*}  
\mathcal{S}_{R}(p_{0},\mathbf{p})&=&{1\over
2}(\gamma^{0}-\mathbf{\gamma}\cdot\hat{p})\,s_{R}(p_{0}, p)\\
 &-&{1\over
2}(\gamma^{0}+\mathbf{\gamma}\cdot\hat{p})\,[s_{R}(-p_{0}^{*}, p)]^{*}.
\end{eqnarray*}
In applications, the fermion propagator appears in combination with 
a fermionic current of the form
\begin{displaymath} \mathcal{J}_{\lambda}(x')=
g \text{Tr}\Big[ e^{-\beta
H}T\Big(A^{\mu}(x')\gamma_{\mu}\overline{\psi}_{\lambda}(x')
\mathcal{O}(y_{1},\dots y_{n})\Big)\Big]{1\over Z},
\end{displaymath}
where $Z=\text{Tr}[\exp(-\beta H)]$ is the partition function.
After projecting out the Dirac matrices from
$\mathcal{S}_{\alpha\lambda}\mathcal{J}_{\lambda}$ 
 a typical application the retarded propagator will produce the
combination
\begin{equation}
\Psi(t,p)=-i\int_{-\infty}^{t}dt'\,s_{R}(t-t',p)\,J(t',p).
\end{equation}
 $\Psi$ is the
response at time $t$ to the source function $J$ acting at an earlier time $t'$.
This can be expressed as an energy integral
\begin{equation}
\Psi(t,p)=\int {dp_{0}\over 2\pi}\,e^{-ip_{0}t}s_{R}(p_{0},p)
\widetilde{J}(p_{0},p),\label{psi}
\end{equation}
where $\widetilde{J}(p_{0},p)$ is the Fourier transform of $J(t,p)$.

\paragraph*{Example:} For the conventional retarded propagator of the
 quark type (with a finite $\gamma$)
the necessary energy integration is
\begin{displaymath}
\Psi(t,p)=\int {dp_{0}\over 2\pi}{e^{-ip_{0}t}\over p_{0}-p+i\gamma/2}
\widetilde{J}(p_{0},p)
\end{displaymath}
If $\widetilde{J}(p_{0},p)$ is analytic in $p_{0}$ in the lower-half plane
and vanishes sufficiently rapidly as $\text{Im}\;p_{0}\to -\infty$, then
the $p_{0}$ contour can be closed in the lower-half plane and the 
result evaluated by Cauchy's theorem.  The simplest way to insure that the source has
these properties in $p_{0}$ is to insist that it vanish for times later than
some $t_{f}$:
\begin{equation}
J(t',p)=\left\{\begin{array}{ll}
0 & t'>t_{f} \\
 \text{arbitrary} & t'<t_{f}.
\end{array}\right.
\end{equation}
The Fourier transform of the source is
\begin{equation}
\widetilde{J}(p_{0},p)=\int_{-\infty}^{t_{f}}dt'\,e^{ip_{0}t'}J(t',p).
\end{equation}
This is analytic for $\text{Im}\; p_{0}<0$ and so the $p_{0}$ integration to be
performed by Cauchy's theorem to give
\begin{equation}
t>t_{f}:\hskip0.2cm \Psi(t,p)=\exp\Big[\big(-ip-{\gamma\over 2}\big)t\Big]
\widetilde{J}(p-i{\gamma\over 2},p).
\end{equation}
The remainder of this section will show that with a current of this type, the
simple function given in Eq. (\ref{ansatz}) produces the analogous result but
with
$\gamma/2$ replaced by $\alpha T[\ln(\omega_{p}t)+C]$.

\subsection{Saddle point integration}

The asymptotic behavior in Eq. (\ref{BI}) was obtained in 
\cite{B1,B2,B3,DB1} by retaining infrared dominant contributions 
in the vicinity $p_{0}\approx p$. The ansatz in Eq. (\ref{ansatz}) is to be
tested in the strip (\ref{strip1}). To isolate the behavior of the propagator
at
$p_{0}\approx p$, it is necessary to introduce some smearing function 
$f(p_{0})$ that suppresses the effects of large $p_{0}-p$.  With the smearing,
Eq. (\ref{psi}) becomes
\begin{displaymath}
\Psi(t,p)=\int{dp_{0}\over 2\pi}e^{-ip_{0} t} s_{R}(p_{0},p)
\widetilde{J}(p_{0},p)f(p_{0}).
\end{displaymath}
Substituting the ansatz Eq. (\ref{ansatz}) gives
\begin{equation}
\Psi(t,p)=-iN\int{dp_{0}\over 2\pi}e^{\phi(p_{0})}
\widetilde{J}(p_{0},p)f(p_{0}),
\end{equation}
where
\begin{equation}
\phi(p_{0})=-ip_{0} t+i\,{p_{0}\!-\!p\over 2\alpha T}
+ a\exp\Big[i\,{p_{0}\!-\!p\over \alpha T}\Big].
\end{equation}
As $t\to\infty$ the integrand oscillates so rapidly  that most values
of $p_{0}$ contribute very little. The  dominant contribution comes
from the region in  which 
$\phi(p_{0})$ is stationary.  
The first derivative of $\phi$ is
\begin{displaymath}
\phi'(p_{0})=-it+i\,{1\over 2\alpha T}
+ {ia\over\alpha T}\exp\Big[i\,{p_{0}\!-\!p\over \alpha T}\Big].
\end{displaymath}
The stationary point $\overline{p}_{0}$  satisfying
$\phi'(\overline{p}_{0})=0$ is
\begin{equation}
\overline{p}_{0}=p-i\alpha T\ln\bigg[{\alpha T t-1/2\over a}\bigg].
\end{equation}
An increasing value of $t$ moves the saddle point further
down into the lower half of the complex plane.
At the saddle point, $\phi$ itself has  the value
\begin{displaymath}
\phi(\overline{p}_{0})\!=-ipt+(\alpha Tt-1/2)\bigg\{-\ln\bigg[\!{\alpha Tt-1/2\over
a}\!\bigg]+1\bigg\}.
\end{displaymath}
 The Taylor series expansion of $\phi(p_{0})$ about the saddle point is
\begin{displaymath}
\phi(p_{0})=\phi(\overline{p}_{0})
+{1\over 2}\phi''(\overline{p}_{0})(p_{0}-\overline{p}_{0})^{2}+\dots
\end{displaymath}
The necessary second derivative is
\begin{displaymath}
\phi''(\overline{p}_{0})=-\bigg[{\alpha Tt-1/2\over(\alpha T)^{2}}\bigg].
\end{displaymath}
The fact that the second derivative is real and
negative means that the value of $\text{Re} \,[\phi(p_{0})]$ is a maximum at the
saddle point and that the value decreases if $p_{0}\!-\!\overline{p}_{0}$ is
positive real or negative real. 
The dominant contribution to the integral is 
\begin{eqnarray*}
\Psi(t,p)&=& -iN\exp[\phi(\overline{p}_{0})]
\int\! {dp_{0}\over 2\pi}\widetilde{J}(p_{0},p)f(p_{0})\\
&&\times \,\exp\bigg\{-{1\over 2}(\alpha Tt-1/2)
\Big[{p_{0}-\overline{p}_{0}
\over \alpha T}\Big]^{2}\bigg\}.
\end{eqnarray*}
The integration contour can be shifted down below the real $p_{0}$ axis.
At large time $\alpha Tt\gg 1$,
if the $p_{0}$ contour is a straight line parallel to the real
axis keeping the difference $p_{0}-\overline{p}_{0}$  real the integrand falls off
very rapidly.  
The integration yields 
\begin{displaymath}
\Psi(t,p)= -i{N\over \sqrt{2\pi}}\exp[\phi(\overline{p}_{0})]
{\alpha T\over \sqrt{\alpha Tt-1/2}}\widetilde{J}(\overline{p}_{0},p)
f(\overline{p}_{0}).
\end{displaymath}
Using the value of $\phi(\overline{p}_{0})$ and $N$ gives
\begin{eqnarray*}
\Psi(t,p)&=&-i\exp\Big\{{-}ipt{-}{1\over 2}{-}\alpha Tt\bigg[\ln\Big[
{\alpha Tt-1/2\over a}\Big]{-}1\bigg]\bigg\}\\
&&\hskip1cm\times \widetilde{J}(\overline{p}_{0},p)f(\overline{p}_{0}).
\end{eqnarray*}
The relation $\ln(\alpha T/a)=\ln(\omega_{p})+C+1$ is useful in simplifying
the result at large time, 
$\alpha Tt\gg 1/2$, and gives
 the final form 
\begin{eqnarray}
\Psi(t,p)&=& -i\exp\Big\{\!-ipt-\alpha Tt
\Big[\ln[\omega_{p}t]+C\Big]\Big\}\nonumber\\
&&\times \widetilde{J}(\overline{p}_{0},p)f(\overline{p}_{0}).
\end{eqnarray}
The time dependence coincides with Eq. (\ref{BI}).
The stationary point is
$\overline{p}_{0}=p-i\alpha T[\ln(\omega_{p}t)+C+1]$ at large times.
This confirms the ansatz.

\section{Soft and collinear photons}

Knowing the electron propagator makes it possible to compute
radiatiave processes.
For a massive electron in a QED plasma there are various infrared divergences
associated with real and virtual photons that eventually cancel. The divergences
arise because if an on-shell electron with $P^{2}=m^{2}$ emits an on-shell
photon with $K^{2}=0$, the electron subsequently propagates with 
amplitude
\begin{equation}
{1\over (P-K)^{2}-m^{2}}={-1\over 2|\mathbf{k}|(E-p\cos\theta)}.
\label{daze}\end{equation} 
For very soft photons, $|\mathbf{k}|\to 0$, this can produce an infrared
divergence.  For example, when an electron of momentum $P$
undergoes a hard scattering to momentum $P'$, infrared
divergences arise from  the real emission and real absorption of soft
photons by the incoming electron and by the outgoing electron. Infrared
divergences also arise from virtual photons in three ways: from electron
self-energy corrections to the ingoing electron and to the outgoing
electron; and from virtual photons that link the incoming and outgoing
electrons (i.e. corrections to the hard scattering vertex). For massive
electrons, the infrared divergences from the real photons exactly cancel
those from the virtual  photons \cite{AW1,Ind}.

If the electron is massless then $E=p$ in Eq. (\ref{daze}) and there would also
be a collinear singularity arising from $1/(1-\cos\theta)$. As Blaizot and Iancu
showed \cite{B2},   the thermal propagator for massless electrons does not
actually have a pole at $p_{0}=p$; it is instead an entire function of
$p_{0}$.
  Since there is no pole, there can be no infrared divergences and no
collinear divergences.

However, even though there are no divergences there is still an
important quantitative question of how  large the radiative corrections are.   
To compute these corrections, the first step is to compute the vertex that 
couples a photon to the electron propagator. At this stage it is convenient 
to return to the strict Bloch-Nordsieck structure of \cite{B1,B2,B3} in which the
 propagating electron has a velocity vector $\hat{v}$. Thus in Eq. (\ref{ansatz}) the
combination
$p_{0}-p$ should be replaced by
 $p_{0}-\hat{v}\cdot\mathbf{p}=P\cdot v$. The Ward-Takahashi identity  for
a photon of four-momentum $K$ to be radiated by an electron with initial
momentum $P$ and final momentum $P-K$ is  
\begin{equation}
-K_{\mu}\Gamma^{\mu}(P-K, P)=s_{R}^{-1}(P-K)-s_{R}^{-1}(P).
\end{equation}
The general solution for the electromagnetic vertex is
\begin{equation}
\Gamma^{\mu}(P-K, P)={v^{\mu}\over K\cdot v}\bigg[
s_{R}^{-1}(P)-s_{R}^{-1}(P-K)\bigg].\label{vertex}
\end{equation}
(In the usual case of a free electron $s_{R}^{-1}(P)=P\cdot v$ so
that $\Gamma^{\mu}(P-K, P)=v^{\mu}$ as appropriate for the Bloch-Nordsieck
approximation.)

Now we employ this in the example of an electron undergoing a hard
scattering from $P$ to $P'$. The initial and final momenta are on
shell;
$P\cdot v=0$ and
$P'\cdot v'=0$. During the hard scattering, either the incoming electron or the
outgoing electron can emit a real photon with four-momentum
$K$.  If $M_{\text{bare}}(P',P)$ is the amplitude for the hard scattering without
radiation, the amplitude for hard scattering with radiation   is approximately
\begin{displaymath}
M_{\text{bare}}(P',P)\epsilon_{\mu}J^{\mu}(K).
\end{displaymath}
 The effective current to which the photon couples is a combination of
the electromagnetic vertex  above and an off-shell electron propagator:
\begin{eqnarray*}
J^{\mu}(K)&=&s_{R}(P-K)\,g\Gamma^{\mu}(P-K,P)\\
&+&g\Gamma^{\mu}(P',P'-K)\,s_{R}(P'-K).
\end{eqnarray*}
Substituting for the vertex functions gives
\begin{equation}
J^{\mu}(K)={gv^{\mu}\over K{\cdot v}}\bigg[{s_{R}(P\!-\!K)\over
s_{R}(P)}-1\bigg] +{gv^{\prime\mu}\over {K\cdot
v}}\bigg[1-{s_{R}(P'\!-\!K)\over s_{R}(P')}\bigg].
\label{current}\end{equation}
When the momentum transfer  $Q^{2}=(P-P')^{2}$ is large, the cross section
for hard scattering with the emission of one real photon ($k_{0}=k$) is
\begin{equation}
\bigg[{d\sigma\over dQ^{2}}\bigg]_{\text{bare}}
\int{d^{3}k\over (2\pi)^{3}2k}(1+n)
\sum_{\text{ pol}}|\epsilon_{\mu}J^{\mu}(K)|^{2},\label{em}
\end{equation}
where $n=1/[\exp(\beta k)-1]$ is the Bose-Einstein function.

For a free electron propagator  $s_{R}^{-1}(P)=0$  and
$s_{R}^{-1}(P')=0$ so that the current is
\begin{displaymath}
J^{\mu}(K)\bigg|_{\text{free}}= {-gv^{\mu}\over
k(1-\hat{k}\cdot
\hat{v})}+{gv^{\prime\mu}\over k(1-\hat{k}\cdot
\hat{v}')}.
\end{displaymath}
This would produce collinear singularities at $\hat{k}\cdot\hat{v}=1$ and
$\hat{k}\cdot\hat{v}'=1$ as well as logarithmic and linear infrared
divergences from
$\int d^{3}k/k^{4}$.

However, since the electron propagator is actually an entire function, it
has no poles. It is a function of the variable ${P\cdot v}$. For small
values of ${K\cdot v}$ the current (\ref{current}) becomes 
\begin{eqnarray}
K\cdot v\ll \alpha T:\hskip0.2cm
J^{\mu}(K)\bigg|_{\text{entire}}\hskip-0.4cm\approx\!\!&& -g{v^{\mu}
\over s_{R}}{ds_{R}\over
dp_{0}}\bigg|_{p_{0}=\mathbf{p}\cdot\hat{v}}\nonumber\\
&&+g{v^{\prime\mu}\over s_{R}}
{ds_{R}\over
dp'_{0}}\bigg|_{p'_{0}=\mathbf{p}'\cdot\hat{v}'}
\end{eqnarray}
In the emission cross section (\ref{em}) there is no collinear divergence.
The integrand becomes $\int d^{3}k/k^{2}$ so there is no infrared
divergence.  The contribution of soft $k$ to  multiplicative factor in Eq.
(\ref{em}) is approximately
\begin{displaymath}
\int^{k_{\text{max}}}{d^{3}k\over (2\pi)^{3}2k}{T\over
k}\bigg({g\over s_{R}}{ds_{R}\over dp_{0}}\bigg)^{2}
={k_{\text{max}}\over \alpha T}{(1+2a)^{2}\over 4\pi}.
\end{displaymath}
Since $k_{\text{max}}\ll \alpha T$ this is a small correction.
A similar estimate applies to the effects produced by the absorption of a
photon and by the exchange of virtual photons.

\section{Discussion}

The approximate propagator in Eq. (\ref{ansatz}) has been shown to 
produce the correct large time behavior in Eq. (\ref{BI}).
This approximate propagator is the first term of the complete expansion of
$s_{R}(p_{0},p)$ that is performed in Appendix B. 
The approximate propagator Eq. (\ref{ansatz}) has some deficiencies.
First, the asymptotic behavior $\ln(\omega_{p}t)$ is only reached at a very
large time
$t$ satisfying 
$\alpha Tt\gg 1$, whereas the asymptotic form in Eq. (\ref{BI})  sets in at
a smaller time 
$\omega_{p}t\gg 1$. Second, the corrections to the asymptotic time dependence
are of order
$1/\alpha Tt$ whereas the corrections to Eq. (\ref{BI}) are  of order
$\exp(-\omega_{p}t)/\omega_{p}t$.

The absence of any singularity in $s_{R}(p_{0},p)$ eliminates infrared
and collinear divergences in radiative processes. However, there are unsettling
consequences in higher orders.  
 For example, the contribution to the
vacuum polarization tensor,
$\Pi^{\mu\nu}(Q)$, of a virtual electron-positron pair would normally have a
branch cut that signals the production of a real $e^{+}e^{-}$ pair.
If $s_{R}(p_{0},p)$ is entire, it appears there can be no
 $e^{+}e^{-}$ production.
This situation arises in zero-temperature models of color  confinement.
There the quark and gluon propagators are thought to have no simple poles
and perhaps no singularities at all. 
A clear example of this  occurs in
  the nonrelativistic quark model. There the confining
potentials used in the Schroedinger equation  produce wavefunctions that 
fall with distance like $\exp[-f(r)]$, where at large distance $f(r)$ grows
more rapidly  than
$r$. Form factors and other matrix elements computed with these
wavefunctions  are entire functions of momentum as a consequence of confinement
\cite{RW}.

\begin{acknowledgments}
This work was supported in part by U.S. National Science Foundation grant
PHY-0099380.
\end{acknowledgments}


\appendix

\section{Complex time and the KMS condition}

The Kubo, Martin, Schwinger
condition \cite{KMS} requires periodicity in complex time of a certain
two point function 
$\mathcal{S}_{>}(x)$ defined below. This   
 section shows that the KMS condition is essentially unrelated to the
behavior of the retarded propagator in complex time.

\subsection{Definitions}

The basic two-point function is
\begin{displaymath}
\big(\mathcal{S}_{>}(x)\big)_{\alpha\beta}=-i\text{Tr}\Big[\varrho\,
\psi_{\alpha}(x)
\overline{\psi}_{\beta}(0)\Big],
\end{displaymath}
where $\varrho=\exp(-\beta H)/\text{Tr}[\exp(-\beta H)]$ is the density operator.
By inserting a complete set of states between the operators, it is easy to
show that this is an  analytic function of time  in the open strip
$-\beta<\text{Im}
\,(x^{0}) <0$.  
The related function 
\begin{displaymath}
\big(\mathcal{S}_{<}(x)\big)_{\alpha\beta}=i\text{Tr}\Big[\varrho\,\overline{\psi}_{\beta}(0) 
\psi_{\alpha}(x)\Big]
\end{displaymath}
is analytic in the open strip $0<\text{Im} \,(x^{0})<\beta$.
The two functions are
related by the Kubo, Martin, Schwinger condition \cite{KMS}
\begin{equation}
\text{Im}\,(x^{0})>0:\;\;
\mathcal{S}_{>}(x^{0}-i\beta,\mathbf{x}
)=-\mathcal{S}_{<}(x^{0},\mathbf{x}).\label{KMS}
\end{equation}

 The retarded and
advanced propagators are related to the thermal average of the
anticommutators:
\begin{eqnarray*}
\mathcal{S}_{R}(x)&=&\theta(x^{0})
\Big[\mathcal{S}_{>}(x)-\mathcal{S}_{<}(x)\Big]\\
\mathcal{S}_{A}(x)&=&\theta(-x^{0})
\Big[-\mathcal{S}_{>}(x)+\mathcal{S}_{<}(x)\Big].
\end{eqnarray*}
The only common region of definition for $\mathcal{S}_{>}(x)$ and 
$\mathcal{S}_{<}(x)$ is for $x^{0}$ real. The retarded and advanced
propagators, with $\theta(\pm x^{0})$ omitted, are usually restricted to 
$x^{0}$ real.

\subsection{Momentum space}

For massless fermions, the combination of chirality and TCP  invariance
makes the propagators a linear combination of $\gamma^{0}$ and
$\vec{\gamma}\cdot\hat{p}$: 
\begin{eqnarray*}
\mathcal{S}_{R}(P)&=&{1\over
2}(\gamma^{0}-\mathbf{\gamma}\cdot\hat{p})\,s_{R}(p_{0},p)\\
&-&{1\over
2}(\gamma^{0}+\mathbf{\gamma}\cdot\hat{p})\,s_{R}^{*}(-p_{0}^{*},p)
\\
\mathcal{S}_{A}(P)&=&{1\over
2}(\gamma^{0}-\mathbf{\gamma}\cdot\hat{p})\,s_{R}^{*}(p_{0}^{*},p)
\\
&-&{1\over
2}(\gamma^{0}+\mathbf{\gamma}\cdot\hat{p})\,s_{R}(-p_{0},p).
\end{eqnarray*}
The propagators are thus determined by  the one function $s_{R}(p_{0},p)$.
Knowing this, one can also construct the two-point functions using
\begin{eqnarray}
\mathcal{S}_{>}(P)
&=&{1\over 1+ \exp(-\beta p_{0})}\Big[\mathcal{S}_{R}(P)
-\mathcal{S}_{A}(P)\Big]\label{S>}\\
\mathcal{S}_{<}(P)
&=&{-1\over 1+ \exp(\beta p_{0})}\;\Big[\mathcal{S}_{R}(P)
-\mathcal{S}_{A}(P)\Big].\label{S<}
\end{eqnarray}
Since the retarded propagator is analytic in the upper-half of the complex $p_{0}$
plane and the advanced is analytic in the lower-half, generally
$\mathcal{S}_{>}(P)$ and $\mathcal{S}_{<}(P)$ are not
analytic anywhere. Their arguments, $p_{0}$, should always be
considered as real.

The Fourier transforms of Eq. (\ref{S>}) and (\ref{S<}) automatically
satisfy the KMS condition (\ref{KMS}). The condition imposes no restriction
on the retarded propagator in momentum space, $\mathcal{S}_{R}(P)$.

\section{Computation of $\mathbf{s_{R}(p_{0},p)}$}

 This  section computes the Fourier transform of Eq. (\ref{BI})
and obtains a complete asymptotic expansion of $s_{R}(p_{0},p)$ valid in
the strip 
\begin{equation}
-{\pi\alpha T\over 2}<\text{Re}(p_{0})-p<{\pi\alpha T\over 2}.\label{strip}
\end{equation} 
The final result is given in Eqs. (\ref{exact}) and (\ref{IJ}).

Before beginning this analysis it is useful to examine the full
 Blaizot and Iancu \cite{B1,B2,B3} result for the
 retarded propagator:
\begin{eqnarray}
s_{R}(p_{0},p)&=&-i\int_{0}^{\infty}\! \!dt\; e^{W_(p_{0},t)}\label{original}\\
W(p_{0},t)&=&i(p_{0}-p)t-\alpha T t \,f(\omega_{p}t)\label{AW1}\\
f(z)=&C&-\gamma+{1-e^{-z}\over z}
+\int_{0}^{z}\!\!ds\,{1-e^{-s}\over s}.\label{f}
\end{eqnarray}
The integral  that defines $f(z)$ can be computed as a
power series: 
\begin{displaymath}
f(z)=C-\gamma+1-
\sum_{k=1}^{\infty}{(-z)^{k}\over k\,
(k+1)!}.
\end{displaymath}
This series converges everywhere in the complex $z$ plane and thus $f(z)$ is an
entire function. In particular, $f(z)$ has no branch cuts. 
However the power series expansion is useless for large $z$.
One can rewrite Eq. (\ref{f}) using repeated integration by parts to
obtain
\begin{eqnarray*}
f(z)&=&\ln(z)+C+{1\over z}
+ {e^{-z}\over z}\sum_{k=1}^{n}{(-1)^{k}k!\over
z^{k}}\\
&+&({-}1)^{n+1}(n{+}1)!\int_{z}^{\infty}\!ds\,{e^{-s}\over
s^{n+2}}.
\end{eqnarray*}
This is also an exact result. Despite the explicit $\ln(z)$, there is no
branch point:
If $z$ is moved in the complex plane
along a counter-clockwise path that encircles the origin, then
$\ln(z)$ changes by $2\pi i$ and the integral over $s$ changes
by $-2\pi i$. Thus $f(e^{2\pi i}z)=f(z)$: there is no
 branch cut. 
For $\text{Re}\,z>0$ and large, one can use the approximate form
\begin{displaymath}
f(z)\approx
\text{ln}(z)+C.
\end{displaymath}

As observed by Blaizot and Iancu \cite{B2}, the integrand in
Eq. (\ref{original}) falls so rapidly at large $t$ (large $z$) that the integral
converges any  complex value of $p_{0}$. Furthermore all the derivatives
$\partial^{n}s_{R}(p_{0},p)/\partial p_{0}^{n}$ are finite at any complex
value of $p_{0}$. Thus $s_{R}(p_{0},p)$ is an entire function of $p_{0}$.

\subsection{Exact computation}

With the simplified integrand, the Fourier transform to be computed is
\begin{equation}
s_{R}(p_{0},p)=-i\int_{0}^{\infty} dt e^{W(p_{0},t)},\label{s1}
\end{equation}
where
\begin{equation}
W(p_{0},t)=i(p_{0}-p)t-\alpha T t\big[\ln(\omega_{p}t)+C\big].\label{W1}
\end{equation}
The computation of the propagator at large $t$ in \cite{B1,B2,B3,DB1}
is most reliable at $p_{0}\approx p$. This will be the range in which the
Fourier transform is performed. 
 It will be necessary to require that
 $p_{0}$ satisfy Eq. (\ref{strip}).

The integration contour in Eq. (\ref{s1}) can be distorted into the complex $t$ plane
and the convergence will not be compromised provided that $\text{Re}\,t\to +\infty$
at the terminus of the contour. Define the complex time
\begin{equation}
\overline{t}={e^{-C-1}\over \omega_{p}}\exp\Big[i\,{p_{0}-p\over \alpha
T}\Big].
\end{equation}
In the complex $t$ plane, the time derivative of $W$
vanishes at $\overline{t}$.
The integration contour in Eq. (\ref{s1}) can be taken as a straight line from the
origin through $\overline{t}$ and since $\text{Re}\,\overline{t}>0$ the integration
converges. The straight line is parametrized by a real variable $u$ such that
$t=\overline{t}\,u$. The exponent becomes
\begin{displaymath}
W(p_{0},t)=\overline{W}\big[-u\ln u+u\big],
\end{displaymath}
where
\begin{equation}
\overline{W}=a\exp\Big[i\,{p_{0}{-}p\over\alpha T}\Big],\label{Wbar}
\end{equation}
and $a$ is given by Eq. (\ref{a}).
It will be convenient to express $W$ as
\begin{displaymath}
W(p_{0},t)= \overline{W}-\overline{W}\big[u\ln u-u+1\big]\\
\end{displaymath} 
The quantity in square brackets vanishes at $u=1$ and is otherwise positive
throughout the domain $0\le u\le \infty$. From $u=0$ to $u=1$ it decreases
monotonically; from $u=1$ to
$u=\infty$ it increases monotonically.

Using $dt=\overline{t}\,du$ and 
$\overline{t}=\overline{W}/\alpha T$, 
the retarded propagator can be expressed as 
\begin{equation}  
s_{R}(p_{0},p)={-i\overline{W}\over\alpha T}
e^{\overline{W}}\;\Big\{I+J\Big\},\label{exact}
\end{equation}
where   the  integral has been split into two parts:  
\begin{eqnarray*}
I&=&\int_{0}^{1}\! \!du\; 
e^{-\overline{W}[u\ln u-u+1]}\\
J&=&\int_{1}^{\infty}\! \!du\; 
e^{-\overline{W}[u\ln u-u+1]}.\label{s2}
\end{eqnarray*}
In the first integral change variables from $u$ to $x$ where
\begin{equation}
0<u<1:\hskip0.5cm{1\over 2}x^{2}=u_{<}\ln(u_{<})-u_{<}+1.\label{x}
\end{equation}
As $u$ increases  from $0$ to $1$, $x$ decreases from $\sqrt{2}$ to $0$. 
Expand $u$ in a series
\begin{displaymath}
u_{<}=1+\sum_{n=1}^{\infty}(-1)^{n}c_{n}x^{n}.
\end{displaymath} 
The first few coefficients are $c_{1}$=1, $c_{2}$=1/6, $c_{3}$=$-$1/72, $c_{4}$=1/270.
The coefficients satisfy the recursion relation 
\begin{displaymath}
\bigg[1-{n\over 2}\bigg]c_{n}=\sum_{k=1}^{n+1}k\, c_{k}\,c_{n+2-k}.
\end{displaymath}
Then $I$ is easily computed
\begin{eqnarray*}
I&=&-\int_{0}^{\sqrt{2}}\!
\!dx\;{du_{<}\over dx}  e^{-\overline{W}x^{2}/2}\\
&=&\sum_{n=1}^{\infty}(-1)^{n+1}nc_{n}\int_{0}^{\sqrt{2}}\!dx\,
x^{n-1}e^{-\overline{W}x^{2}/2}.
\end{eqnarray*}
In the second integral change variables to $y$ where
\begin{equation}
1<u<\infty:\hskip0.5cm{1\over 2}y^{2}=u_{>}\ln(u_{>})-u_{>}+1.\label{y}
\end{equation}
As $u$ increases from $1$ to $\infty$, $y$ increases from $0$ to $\infty$. 
Expand $u$ in a series
\begin{displaymath}
u_{>}=1+\sum_{n=1}^{\infty}c_{n}y^{n},
\end{displaymath}
with the same $c_{n}$. Then $J$ is
\begin{eqnarray*}
J&=&\int_{0}^{\infty}\!
\!dy\;{du_{>}\over dy}  e^{-\overline{W}y^{2}/2}\\
&=&\sum_{n=1}^{\infty}nc_{n}\int_{0}^{\infty}\!dy\, y^{n-1}e^{-\overline{W}y^{2}/2}
.\end{eqnarray*}
In the sum $I+J$ there is a partial cancellation. The integrations can be
performed exactly with the result 
\begin{eqnarray}
&&I+J=\nonumber\\
&&\sum_{k=0}^{\infty}(2k+1)c_{2k+1}\bigg({-2}{d\over
d\overline{W}}\bigg)^{k}
\sqrt{{\pi\over
2\overline{W}}}\,\bigg[2+\text{erfc}\Big(\sqrt{\overline{W}}\Big)\bigg]\nonumber\\
&&+\sum_{k=0}^{\infty}(2k+2)c_{2k+2}\bigg({-2}{d\over d\overline{W}}\bigg)^{k}
{e^{-\overline{W}}\over\overline{W}}.\label{IJ}
\end{eqnarray}
The asymptotic behavior of the complementary error function is
\begin{displaymath}
\text{erfc}\Big(\!\sqrt{\overline{W}}\,\Big)={e^{-\overline{W}}\over
\sqrt{\pi\overline{W}}}\bigg[1-{1\over 2\overline{W}}
+{3\over 4\overline{W}^{2}}+\dots \bigg]
\end{displaymath}
 Substitution of Eq. (\ref{IJ}) into Eq. (\ref{exact}) gives a complete  asymptotic
expansion for the retarded propagator when $p_{0}$ satisfies (\ref{strip}). 
Although this expansion is periodic under $p_{0}\to p_{0}+4\pi\alpha T$,  
such a shift is well outside the domain of validity given in Eq. (\ref{strip}).

\subsection{Large $\overline{W}$ limit}

The asymptotic series (\ref{IJ})  obviously behaves very badly at small
values  of $\overline{W}$. The most useful application of the results is clearly when
$\text{Re}\, \overline{W}^{1/2}$ is large and positive. 
In this region  the exponentials are negligible
\begin{eqnarray}
s_{R}(p_{0},p)&=&-i{\sqrt{2\pi\overline{W}}\over\alpha
T}e^{\overline{W}}\sum_{k=0}^{\infty}c_{2k+1}
{(2k+1)!!\over\overline{W}^{k}}\label{powers}\\
 &=&-i{\sqrt{2\pi\overline{W}}\over\alpha
T}e^{\overline{W}}\bigg[1-{24\over \overline{W}} -{23\over 1152
\overline{W}^{2}}+\dots \bigg].\nonumber
\end{eqnarray}

The simplest approximation is to keep only the first term
in Eq. (\ref{powers}).
In terms of $p_{0}$ and $p$ the propagator is
\begin{equation}
 s_{R}(p_{0},p)=-i{\sqrt{2\pi a}\over\alpha
T}\exp\bigg[i\,{p_{0}\!-\!p\over 2\alpha T} + a\exp\Big[i\,{p_{0}\!-\!p\over
\alpha T}\Big]\bigg].
\end{equation}
This is
 analytic for any finite $p_{0}$;  it vanishes in the limit
$\text{Im}\;p_{0}\to +\infty$; and it diverges when $\text{Re}\,p_{0}=p$ and
$\text{Im}\;p_{0}\to -\infty$ in agreement with Eq. (\ref{bound}). This is the form 
used throughout the paper.

Before concluding, it is necessary to discuss how $\overline{W}$ can be large.
From Eq. (\ref{Wbar}) the absolute value is  
\begin{equation}
\big|\overline{W}\big|=ae^{-\text{Im}\,p_{0}/\alpha T}={3g\over 4\pi}
\exp\Big[-C-1-{\text{Im}\,p_{0}\over\alpha T}\Big]
\end{equation}
For  small $g$ there are two ways in which the magnitude  $|\overline{W}|$ can be large.
One possibility is to restrict consideration to the region in which
$\text{Im}\,p_{0}/\alpha T$ is large and negative. This is perfectly valid, but it
excludes the physically important region of real $p_{0}$. The
other possibility is for the constant $C$ to be large and negative.

 Both groups \cite{B3,DB1} showed that value of $C$ in the  Coulomb gauge is 
$C=\ln\sqrt{3}+\gamma_{E}-1=0.12652\dots$
In covariant gauges the value of $C$ can be different. 
When the
  zero temperature, Feynman propagator for the photon is
\begin{displaymath}
D^{\mu\nu}_{F}(q)={1\over q^{2}+i\epsilon}
\bigg[-g^{\mu\nu}+(1-\lambda){q^{\mu}q^{\nu}\over q^{2}}\bigg],
\end{displaymath}
it is necessary to impose an
infrared cutoff $\mu$.
Blaizot and Iancu \cite{B3} discuss two limits: $\mu t\ll 1$ and $\mu t\gg 1$. 
If $\mu t\gg 1$ then $C$ is independent of the gauge parameter $\lambda$.
However in the calculation of Sec. B1, the dominant contribution comes from
finite values of $t$. (The integral
$I$ sums  $t$ in the range $0\le t\le \overline{t}$. If the integral limits on $J$
are changed to $0\le y\le \sqrt{2}$, corresponding to $\overline{t}\le t\le
e\overline{t}$, the error made is $2.35\times e^{-\overline{W}}/\overline{W}$,
which is negligible.) Since the important values of $t$ are finite in the previous
integration, then as the infrared cutoff $\mu$  is reduced
 the appropriate limit is $\mu t\ll 1$. In this limit Blaizot and Iancu \cite{B3}
showed that
$C$ does depend on the gauge parameter $\lambda$:
\begin{equation}
C={\lambda \over 2}+\ln\sqrt{3}+\gamma_{E}-1={\lambda\over 2}+0.12652\dots
\end{equation}
Consequently a large, negative $\lambda$ will make $C$ large and negative, which will make
$|\overline{W}|$ large thus justifying the approximation. In \cite{KS} the range
$-857<\lambda<433$ was explored.

\end{document}